\documentstyle[floats,epsf,twocolumn,prb,aps]{revtex}

\begin{document}
\draft 

\wideabs{

\title{Doping dependent evolution of the electronic
structure of La$_{2-x}$Sr$_x$CuO$_4$\\ in the 
superconducting and metallic phases}

\author{A.~Ino$^1$\cite{adr1}, C.~Kim$^2$, M.~Nakamura$^3$,
T.~Yoshida$^1$, T.~Mizokawa$^1$, A.~Fujimori$^1$, Z.-X.~Shen$^2$,\\
T.~Kakeshita$^4$, H.~Eisaki$^4$ and S.~Uchida$^4$}

\address{$^1$ Department of Physics and Department of 
Complexity Science and Engineering,\\ University of 
Tokyo, Bunkyo-ku, Tokyo 113-0033, Japan} 

\address{$^2$ Department of Applied Physics and Stanford Synchrotron 
Radiation Laboratory,\\ Stanford University, Stanford, CA94305, USA}

\address{$^3$ Department of Physics, Nara University of Education, 
Takabatake-cho, Nara 630-8528, Japan}

\address{$^4$ Department of Advanced Materials Science, University of Tokyo, 
Bunkyo-ku, Tokyo 113-8656, Japan}

\date{\today} 

\maketitle
\begin{abstract}
	The electronic structure of the La$_{2-x}$Sr$_x$CuO$_4$ (LSCO) 
	system has been studied by angle-resolved photoemission 
	spectroscopy (ARPES).  We report on the evolution of the Fermi 
	surface, the superconducting gap and the band dispersion 
	around the extended saddle point ${\bf k}=(\pi,0)$ with hole 
	doping in the superconducting and metallic phases.  As hole 
	concentration $x$ decreases, the flat band at $(\pi,0)$ moves 
	from above the Fermi level ($E_{\rm F}$) for $x>0.2$ to below 
	$E_{\rm F}$ for $x<0.2$, and is further lowered down to 
	$x=0.05$.  From the leading-edge shift of ARPES spectra, the 
	magnitude of the superconducting gap around $(\pi,0)$ is found 
	to monotonically increase as $x$ decreases from $x=0.30$ down 
	to $x=0.05$ even though $T_c$ decreases in the underdoped 
	region, and the superconducting gap appears to smoothly evolve 
	into the normal-state gap at $x=0.05$.  It is shown that the 
	energy scales characterizing these low-energy structures have 
	similar doping dependences.  For the heavily overdoped sample 
	($x=0.30$), the band dispersion and the ARPES spectral 
	lineshape are analyzed using a simple phenomenological 
	self-energy form, and the electronic effective mass 
	enhancement factor $m^*/m_b \simeq 2$ has been found.  As the 
	hole concentration decreases, an incoherent component that 
	cannot be described within the simple self-energy analysis 
	grows intense in the high-energy tail of the ARPES peak.  Some 
	unusual features of the electronic structure observed for the 
	underdoped region ($x \lesssim 0.10$) are consistent with the 
	numerical works on the stripe model.
\end{abstract}

}

\narrowtext

\section{Introduction}

For the detailed understanding of a high-$T_c$ cuprate
system, the determination of the low-energy electronic
structure, i.e., the Fermi surface, the band dispersion and
the superconducting and normal-state gaps, is required as
the ground for studies of superconducting mechanism and for
the interpretation of thermodynamic and transport
properties.  Indeed, such information has been directly
observed by angle-resolved photoemission spectroscopy
(ARPES) for Bi$_2$Sr$_2$CaCu$_2$O$_{8+y}$
(Bi2212),\cite{Dessau-rev,Marshall,DingDoping,Ding-dgap,DingGap,PJWhite,Loeser,Harris,Norman}
Bi$_2$Sr$_2$CuO$_{6+y}$
(Bi2201)\cite{Harris-Bi2201,Bi2201-FS} and
YBa$_2$Cu$_3$O$_{7-y}$ (YBCO).\cite{YBCO} Since the
electronic properties of the high-$T_c$ cuprates are
strongly dependent on the hole concentration, it is
necessary to investigate the doping dependence of ARPES
spectra systematically over a wide hole concentration range
in order to extract key features relevant to the high-$T_c$
superconductivity.

Among the high-$T_c$ cuprate systems, we have recently
focused on the La$_{2-x}$Sr$_x$CuO$_4$ (LSCO) system
\cite{FermiSurface,twocomponents} because the hole
concentration is well controlled over an exceptionally wide
range and uniquely determined by the Sr concentration $x$
(and small oxygen non-stoichiometry).  In addition, an
instability towards spin-charge ordering in a stripe form
has been extensively discussed from the incommensurate
inelastic neutron peaks.\cite{Tranquada,Zaanen,Emery} The
suppression of $T_c$ at $x\sim1/8$ \cite{LBCO-1/8,LNSCO-1/8}
indicates that the stripe fluctuation has more static
tendency in LSCO than in Bi2212.

In this paper, we address the evolution of the Fermi surface, the 
superconducting gap and the band dispersions with hole doping 
throughout the superconducting and metallic phases ($0.05 \le x \le 
0.30$) of LSCO, focusing on the features around the extended saddle 
point at ${\bf k}=(\pi,0)$, which are crucial to the determination of 
the Fermi surface topology and the behaviors of superconducting and 
normal-state gaps.  The discussion leads to the issue of the doping 
dependence common to three characteristic energies of the electronic 
structure, and the self energy and the electron effective mass are 
deduced.  In the previous paper, ARPES spectra for $x=0.10$ and $0.30$ 
have been reported and the formation of a Fermi surface centered at 
$(0,0)$ for an overdoped sample has been addressed.\cite{FermiSurface} 
On the other hand, the evolution of the ARPES spectra around the 
superconductor-insulator transition ($x\simeq 0.05$) has been 
addressed in Ref.~\onlinecite{twocomponents}, where the suppression of 
quasiparticle weight around $(\pi/2,\pi/2)$ has been also discussed 
for underdoped superconducting LSCO.\cite{twocomponents}

\section{Experimental}

Single crystals of La$_{2-x}$Sr$_x$CuO$_4$ were grown by the
traveling-solvent floating-zone method and were annealed so
that the oxygen content became stoichiometric.  The accuracy
of the hole concentration was $\pm0.01$.  The samples were
insulating for $x=0.05$, superconducting for $x=0.10$, 0.15
and 0.22, and metallic without superconductivity for
$x=0.30$.  Details of the growth conditions and
characterization of the crystals are described
elsewhere.\cite{Nakamura&Uchida,Uchida&Tamasaku,Tamasaku1}

ARPES measurements were carried out at the undulator beamline~5-3 of 
Stanford Synchrotron Radiation Laboratory (SSRL).  Incident photons 
had energies of $h\nu=29$ or 22.4 eV and were linearly polarized.  The 
electric vector and the wave vector of the incident photons and the 
sample surface normal were kept in the horizontal plane.  The samples 
were fixed with respect to the incident light with an incident angle 
of 45$^\circ$ and ARPES spectra were collected using a hemispherical 
analyzer of 50 mm radius.  The total instrumental resolution including 
the analyzer and the monochromator was approximately 45 meV and the 
angular acceptance was $\sim \pm 1^\circ\!$.  In the case of LSCO, 
$1^\circ$ corresponds to 1/19 and 1/23 of the $(0,0)-(\pi,0)$ distance 
in the Brillouin zone (BZ) of the CuO$_2$ plane for the incident 
photon energies of $h\nu=29$ and 22.4 eV, respectively.  The samples 
were cleaved {\it in situ} at the plane parallel to the CuO$_2$ planes 
by knocking a top-post glued on the sample under an ultra high vacuum 
better than $5\times 10^{-11}$ Torr.  The orientation of the sample 
surface normal was finely readjusted using the reflection of a laser 
beam.  The direction of the $a$- and $b$-axes were finely corrected 
using the band folding in the ARPES spectra with respect to the 
$k_{y}=0$ line.  Since the sample surface degraded rapidly at high 
temperatures, the samples were kept at low temperatures ($T\simeq 11$ 
K) during the measurements.  The cleanliness of the surface was 
checked by the absence of a hump at energy $\sim -9.5$ eV and of a 
shoulder of the valence band at $\sim -5$ eV. All the spectra 
presented here were taken within 12 hours after cleaving.  The 
position of the Fermi level ($E_{\rm F}$) was repeatedly calibrated 
with gold spectra during the measurement and the experimental 
uncertainty in the energy calibration was about $\pm 2$ meV. The 
intensities of the spectra at different angles have been normalized to 
the intensity of the incident light.  In the present paper, the 
measured crystal momenta ${\bf k} = (k_x,k_y)$ are referred to in 
units of $1/a$, where $a$ is twice the Cu-O bond length within the 
CuO$_2$ plane, and the extended zone notation is adopted, that is, a 
$k_x$ value larger than $\pi$ means that the momentum is in the second 
BZ.

\section{Results}

\subsection{ARPES spectra}

\begin{figure}[!b]
	\epsfxsize=84mm \centerline{\epsfbox{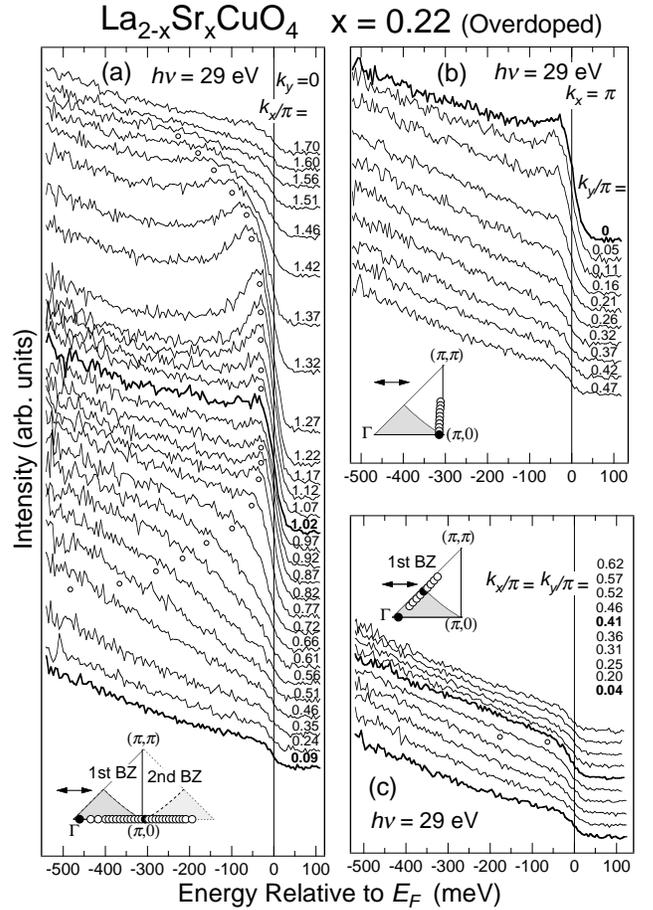}}
	\vspace{1pc} \caption{ARPES spectra of overdoped
	La$_{2-x}$Sr$_x$CuO$_4$ ($x=0.22$) without any data
	manipulations except for the energy calibration. 
	Insets show the measured momenta (circles) in the
	Brillouin zone and the in-plane component of the
	polarization of the incident photons (arrows).  In
	going along $(0,0)\rightarrow(\pi,0)$, the band
	crosses $E_{\rm F}$ near $(\pi,0)$, although part of
	spectral weight remains below $E_{\rm F}$ at
	$(\pi,0)$.}
\label{x=0.22}
\end{figure}

ARPES spectra of overdoped LSCO ($x=0.22$) in the superconducting 
state are shown in Fig.~\ref{x=0.22}.  Here, the spectra are the raw 
data recorded on the spectrometer except for that the energies have 
been calibrated to the Fermi edge of gold.  It seems that the 
intensity of the dispersive component relative to the 
angle-independent background is weaker for LSCO than that for Bi2212.  
Probably, since the cleaved surface of LSCO is not so flat as that of 
Bi2212, some photoelectrons lose the momentum information at the 
surface of LSCO and thus detected as an angle-independent background.  
In addition, the peak intensity is also strongly affected by the 
transition matrix element, which is different among various cuprate 
materials.  The relative weakness of the dispersive component due to the 
high background may induce some uncertainty of the spectral lineshape, 
compared to Bi2212.  However, the peak energy is less affected by it, 
and the peak width of LSCO is practically similar to that of Bi2212 
under the similar doping level and the same instrumental resolution 
\cite{Dessau}.  Indeed, the energy position and width of the peak were 
well reproduced by several experiments, indicating validity of the 
analysis of the ARPES peak performed in Section III.E.

Usually the band dispersion is obtained by tracing the ARPES spectral 
peak.  As one goes from (0,0) to $(\pi,0)$ or from $(2\pi,0)$ to 
$(\pi,0)$, the peak energy increases towards $E_{\rm F}$ as shown in 
Fig.~\ref{x=0.22}~(a).  Around $\sim(0.8\pi,0)$ and $\sim(1.2\pi,0)$, 
the peak reaches the vicinity of the Fermi level ($E_{\rm F}$) and the 
peak intensity decreases between these points.  However, part of the 
spectral weight remains below $E_{\rm F}$ even at $(\pi,0)$, and the 
weight completely disappears only in going from $(\pi,0)$ to 
$(\pi,\pi)$ [Fig.~\ref{x=0.22}~(b)].  The remnant weight at $(\pi,0)$ 
is larger for $x=0.22$ than for $x=0.3$,\cite{FermiSurface} indicating 
that a band of flat dispersion around $(\pi,0)$ lies quite close to 
the Fermi level for $x=0.22$.\cite{Bi2201-FS,Markiewicz} Since even 
for $x=0.3$ small weight remains below $E_{\rm F}$ at 
$(\pi,0)$,\cite{FermiSurface} the band around $(\pi,0)$ is not a 
single peak but has a broad energy distribution, implying a 
complicated spectral weight distribution around $(\pi,0)$ as discussed 
recently.\cite{Chuang,Fretwell} Along the 
(0,0)$\rightarrow$$(\pi,\pi)$ cut, although the dispersive feature is 
weak, the increase of the intensity at $E_{\rm F}$ compared to the 
background around $(0.4\pi,0.4\pi)$ suggests a Fermi-surface crossing 
as in $x=0.3$ and 0.15.\cite{FermiSurface,twocomponents} Overall, the 
electronic structure for $x=0.22$ is in transition between the 
electronic structures characterized by the Fermi surfaces centered at 
$(0,0)$ ($x=0.30$) and at $(\pi,\pi)$ ($x=0.15$).\cite{FermiSurface}

\begin{figure}[!t]
	\epsfxsize=84mm \centerline{\epsfbox{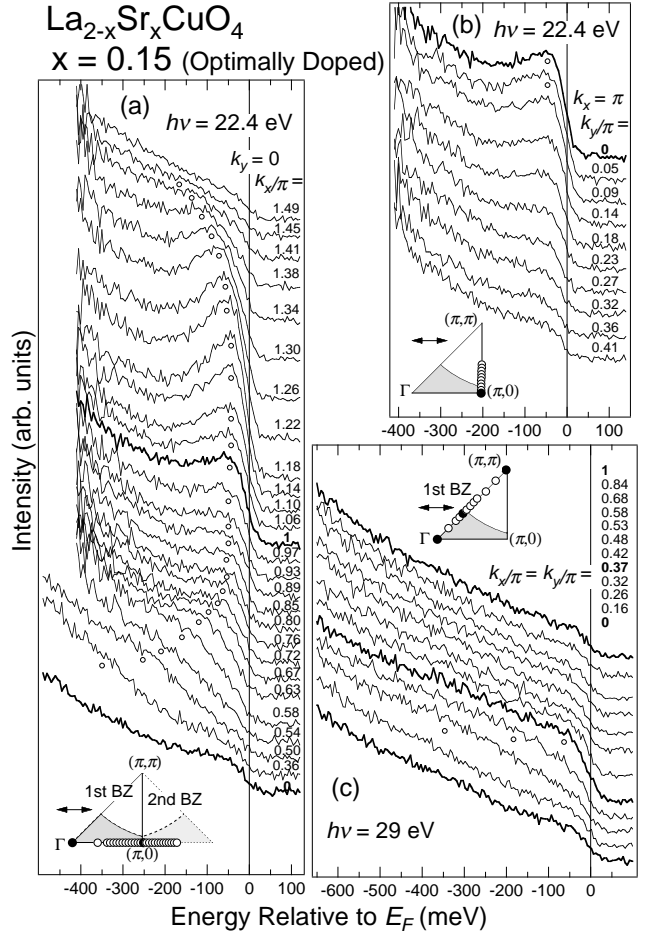}}
	\vspace{1pc} \caption{ARPES spectra of optimally
	doped La$_{2-x}$Sr$_x$CuO$_4$
	($x=0.15$),\protect\cite{FermiSurface,twocomponents}
	displayed in a similar way to Fig.~\ref{x=0.22}. 
	Along $(0,0)\rightarrow(\pi,0)$, the peak clearly
	remains below $E_{\rm F}$ , indicating a Fermi
	surface centered at $(\pi,\pi)$.  The band around
	$(\pi,0)$ shows a very flat dispersion and is
	located slightly below $E_{\rm F}$.}
\label{x=0.15}
\end{figure}

In Fig.~\ref{x=0.15}, ARPES spectra for optimally doped LSCO
($x=0.15$) are displayed
again\cite{FermiSurface,twocomponents} in a similar way to
Fig.~\ref{x=0.22}.  Even though the spectra were taken at a
temperature ($\sim 11$ K) well below $T_c$ ($\simeq39$ K),
the condensation peak is absent or unresolved for LSCO as in
Bi2201,\cite{Harris-Bi2201} while the lineshape with a peak,
dip and hump has been observed around $(\pi,0)$ for
Bi2212.\cite{Dessau-rev,LoeserPeak,NormanPeak} As one goes
from $(0,0)$ to $(\pi,0)$ or from $(2\pi,0)$ to $(\pi,0)$,
the peak approaches $E_{\rm F}$ but clearly remains below
$E_{\rm F}$ at $(\pi,0)$, indicating a Fermi surface
centered at $(\pi,\pi)$.  In going from $(\pi,0)$ to
$(\pi,\pi)$, the peak intensity decreases, while the
midpoint of a leading-edge is always below $E_{\rm F}$ ($-3$
meV at the closest to $E_{\rm F}$, i.e., the minimum-gap
locus), implying that the band goes above $E_{\rm F}$
through the superconducting gap.  The band around $(\pi,0)$
shows a very flat dispersion and is located slightly below
$E_{\rm F}$.\cite{Bi2201-FS,Markiewicz} The spectra along
$(0,0)\rightarrow(\pi,\pi)$ for $x=0.15$ are similar to
those for $x=0.3$\cite{FermiSurface} and 0.22: one can
identify the dispersion of the weak feature crossing $E_{\rm
F}$ at $\sim(0.4\pi,0.4\pi)$.  Thus, the electronic
structure for $x=0.15$ is similar to those for other
optimally doped Bi2212\cite{Marshall} and
Bi2201,\cite{Bi2201-FS} except for that the dispersive
spectral peak along $(0,0)\rightarrow(\pi,\pi)$ is weak for
LSCO.

\begin{figure}[!t]
	\epsfxsize=84mm \centerline{\epsfbox{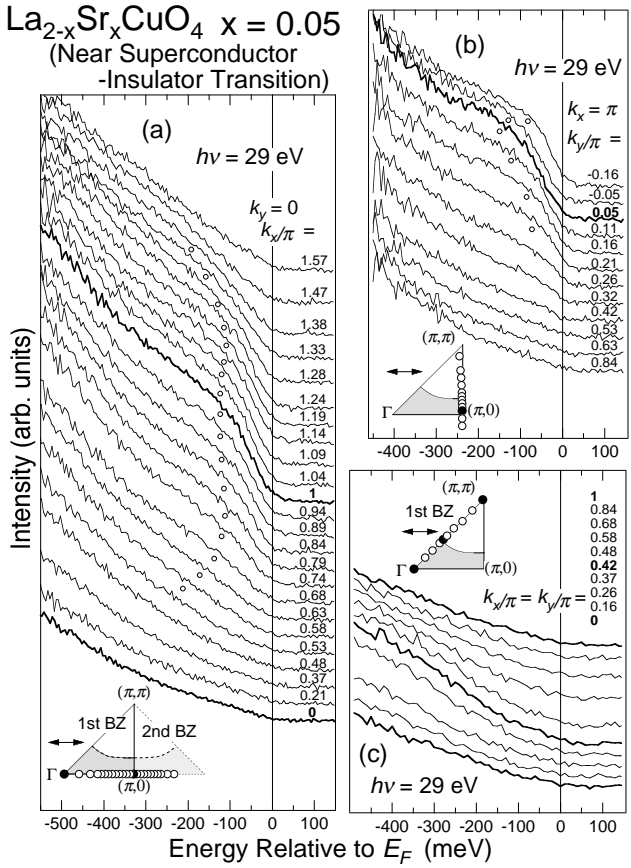}}
	\vspace{1pc} \caption{ARPES spectra near the Fermi
	level ($E_{\rm F}$) for La$_{2-x}$Sr$_x$CuO$_4$
	($x=0.05$) near the superconductor-insulator
	transition, displayed in a similar way to
	Fig.~\ref{x=0.22}.  Along
	$(\pi,0)\rightarrow(\pi,\pi)$, the peak disperses
	towards $E_{\rm F}$ and loses its intensity before
	reaching $E_{\rm F}$ around $(\pi,0.25\pi)$,
	indicating a ``normal-state gap'' opened on the
	underlying Fermi surface.}
\label{x=0.05}
\end{figure}

ARPES spectra of the heavily underdoped LSCO ($x=0.05$) in
the normal state are shown in Fig.~\ref{x=0.05}.  As the
hole concentration $x$ decreases, the peak near $E_{\rm F}$
around $(\pi,0)$ becomes broader and weaker. 
This is consistent with the spectra of other underdoped cuprates,
where the dispersive feature is so broad
that it is merely a shoulder rather
than a spectral peak.\cite{DingDoping,DingGap}
When the hole concentration $x$ decreases down to $x \leq
0.03$ for LSCO, the feature near $E_{\rm F}$ becomes too weak to
discuss the dispersion because of the spectral weight
transfer into a band around $-0.5$ eV (see
Fig.~\ref{disparound}).\cite{twocomponents} As shown in
Fig.~\ref{x=0.05}, while the band for $x=0.05$ stays below
$E_{\rm F}$ with very weak dispersion along
$(0.8\pi,0)\rightarrow(\pi,0)$, the band disperses
rather strongly towards $E_{\rm F}$ along
$(\pi,0)\rightarrow(\pi,0.2\pi)$ and the feature
disappears around $(\pi,0.25\pi)$ before the leading-edge
midpoint reaches $E_{\rm F}$, indicating that a gap is
opened around $(\pi,0.25\pi)$ for $x=0.05$.  Presumably,
the gap is opened on the underlying Fermi surface as in the
superconducting samples although $T_c\simeq0$, and may be
regarded as a ``normal-state
gap''.\cite{DingGap,PJWhite,Loeser,Harris,Norman}
Remarkably, in the $(0,0)\rightarrow(\pi,\pi)$ cut, no
dispersive feature nor intensity modulation could be
identified at $\sim$$E_{\rm F}$ for $x\leq0.12$
\cite{twocomponents} in contrast to the spectra for
$x\geq0.15$.  Therefore, the electronic structure near
$E_{\rm F}$ for $x=0.05$ is similar to that for $x=0.1$
reported in the previous paper \cite{FermiSurface}: the 
Fermi surface centered at $(\pi,\pi)$ is observed around
$(\pi,0.25\pi)$, but it is invisible around $(\pi/2,\pi/2)$.

\subsection{Band dispersions}

\begin{figure}[!t]
	\epsfxsize=84mm \centerline{\epsfbox{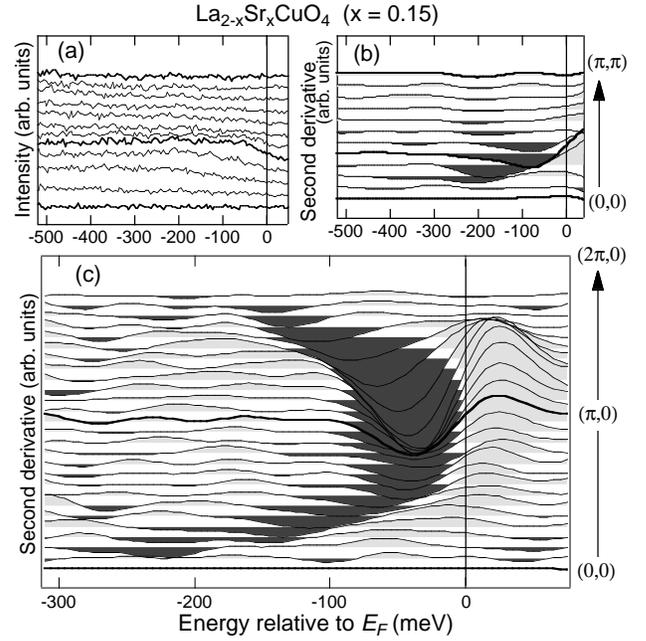}} 
	\vspace{1pc} \caption{(a) Dispersive component of ARPES 
	spectra taken along $(0,0)\rightarrow(\pi,\pi)$ for $x=0.15$.  
	The angle-independent background has been subtracted 
	from the original spectra shown in Fig.~\ref{x=0.15} (c).  (b) 
	and (c) Second derivatives of the ARPES spectra taken along 
	$(0,0)\rightarrow(\pi,\pi)$ and 
	$(0,0)\rightarrow(\pi,0)\rightarrow(2\pi,0)$, respectively, 
	for $x=0.15$.  The positive and negative peaks in the second 
	derivatives are colored with light and dark gray, 
	respectively, in the figure.  Original ARPES spectra are shown 
	in Fig.~\ref{x=0.15}.  The dispersive component of the ARPES 
	spectra has been smoothed by convoluting with the gaussian, 
	and then differentiated twice (see the text).  The results 
	are displayed in Figs.~\ref{flatband} and 
	\ref{disparound} in order to visualize the dispersion 
	relation.}
\label{secdev}
\end{figure}

\begin{figure}[!t]
	\epsfxsize=84mm \centerline{\epsfbox{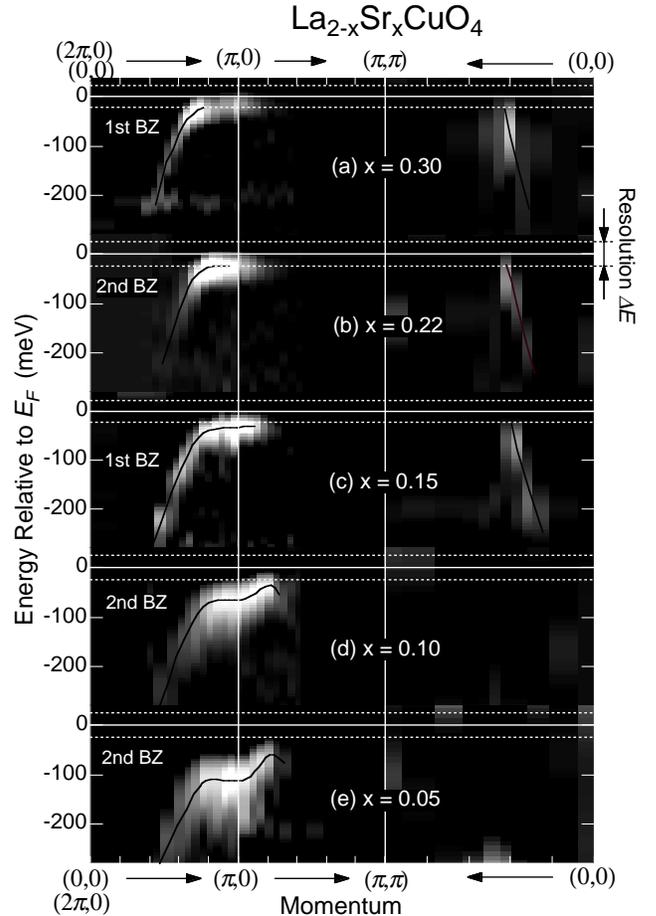}}
	\vspace{1pc}
	\caption{Band dispersion near the Fermi level for
	La$_{2-x}$Sr$_x$CuO$_4$, measured by ARPES. The second
	derivatives of the ARPES spectra,
	which are shown in Fig.~\ref{secdev} for example,
	are displayed as a density plot on the gray scale, where 
	white regions denote the negative peak of the second derivatives.
	Note that, because
	of the finite instrumental resolution $\Delta E \sim 45$
	meV, the structure near $E_{\rm F}$ are pushed down
	below the resolution limit $\sim - \Delta E/2$ (dashed
	lines).}
\label{flatband}
\end{figure}

Overall band dispersions near $E_{\rm F}$ are visualized in 
Fig.~\ref{flatband} by use of the second derivatives, which are shown 
in Fig.~\ref{secdev} for example.  First, the step at $E_F$ seen in 
the spectrum at (0,0) seems to be present at all the angles with 
almost constant intensity, as shown in Figs.~\ref{x=0.22}, 
\ref{x=0.15} and \ref{x=0.05}.  Hence, we assigned the spectrum at 
(0,0) to the angle-integrated signals likely due to the surface 
imperfection, because no emissions are allowed at (0,0) from the 
$d_{x^2-y^2}$ symmetry of Cu 3$d$ orbitals due to the photoemission 
matrix-element effect.  In order just to remove this extrinsic step, 
the spectra at (0,0) was subtracted from all the spectra at the other 
angles under simple normalization to the intensity of the incident 
light.  The validity of this subtraction may be understood by a 
typical result shown in Fig.~\ref{secdev}(a).  Indeed, the spectrum at 
(0,0) is so featureless that its subtraction makes essentially no 
effect on the second derivatives except for the extrinsic step at 
$E_F$.  All the resulting spectra were then smoothed by convoluting 
with the gaussian whose energy width is the order of the energy 
resolution (typically $\sim 50$ meV), since the collected signals were 
of the order of $\sim 10^3$ counts for the peak component and thus the 
signal to noise ratio is the order of $\sim 1/30$.  Along the momentum 
direction, no smoothing nor interpolation is applied to the data and 
thus each horizontal pixel in Fig.~\ref{flatband} corresponds to each 
ARPES spectrum.  Finally, the spectra are differentiated two times and 
displayed by the gray scale plot in Fig.~\ref{flatband}, where white 
regions denote the negative peak of the second derivatives.  In the 
differentiation, the energy step of the data was smaller enough (5 or 
10 meV) compared to the energy resolution.  Indeed, taking the second 
derivatives would be an appropriate way to visualize the band 
dispersion of this system, because for $x \lesssim 0.1$ the dispersive 
feature does not show a clear peak but a shoulder.  Practically, the 
second derivative method has been widely used and outlined the band 
dispersions excellently from the ARPES 
spectra.\cite{second1,second2,second3,second4,second5,second6,second7} 
The validity of the above data manipulations is assured by comparing 
the second derivatives in Fig.~\ref{secdev} (c) with the original raw 
spectra in Fig.~\ref{x=0.15} (a), and comparing Fig.~\ref{flatband} 
with the gray scale plot of the original data shown in the top panels 
of Fig.~\ref{selfenergy} for $x=0.30$ and 0.15.  In 
Fig.~\ref{flatband}, thin black curves following the negative peaks in 
the second derivatives are also drawn.  Thus their error bars were 
represented by the half width of the white gradation.  Note that, 
because of the Fermi cut-off and the finite instrumental resolution 
$\Delta E \sim 45$ meV, spectral features near $E_{\rm F}$ are pushed 
down below $\sim - \Delta E/2$ (dashed lines).\cite{broadening} The 
obtained band dispersion for $x=0.15$ is similar to the ARPES results 
of other optimally doped cuprates such as Bi2212\cite{Marshall} and 
Bi2201.\cite{Bi2201-FS}

Figure~\ref{flatband} shows that the so-called ``$(\pi,0)$
flat band'' is clearly observed for $x=0.15, 0.10$ and 0.05
in the sense that the flat region around the saddle point at
$(\pi,0)$ is extended up to $\sim(0.7\pi,0)$.\cite
{Bi2201-FS,Markiewicz} The flat band, which is $\sim 120$
meV below $E_{\rm F}$ for $x=0.05$, moves upwards
monotonically with hole doping, crosses the Fermi level
around $x\simeq0.2$ causing the increase of the density of
states (DOS) at $E_{\rm F}$ as observed by angle-integrated
photoemission (AIPES) \cite{AIPES} and the quasiparticle
density reflected in the electronic specific
heat,\cite{Momono} and finally goes above the Fermi level. 
Since the chemical potential shift with hole doping is small
($\ll 100$ meV) in the region $0\leq x\leq 0.15$,\cite{chem}
the energy shift of the flat band in this composition range
is due to the deformation of the band structure itself. 
Probably the lowering of the flat band at $(\pi,0)$ is due
to the influence of short-range antiferromagnetic
correlations.  Under the antiferromagnetic correlations, the
spectral function of magnetic excitations
$\chi''(\mathbf{q},\omega)$ is peaked near
$\mathbf{q}=(\pi,\pi)$.  Then, the photohole at $(\pi,0)$ is
particularly dressed strongly in the collective magnetic
excitations, because the photohole at $(\pi,0)$ can enter in
the state of similar energy around $(0,\pi)$ with producing
a collective excitation $\mathbf{q}=(\pi,\pi)$, as proposed
by Shen and Schrieffer.\cite{Shen&Schrieffer,Kim} Since the
emissions from the dressed photoholes are predominant among
the spectral intensity in the underdoped region, the kinetic
energies of photoelectrons from $(\pi,0)$ are lowered by the
stronger dressing of photoholes with decreasing hole
concentration.


As for the underdoped samples ($x=0.05$ and 0.10), the band
dispersion around $(\pi,0)$ is not symmetric between the
$(\pi,0)$$\rightarrow$$(0,0)$ and
$(\pi,0)$$\rightarrow$$(\pi,\pi)$ directions.  While the
band is very flat showing almost no dispersion along
$(\pi,0)$$\rightarrow$$(0.7\pi,0)$, the dispersion along
$(\pi,0)$$\rightarrow$$(\pi,0.3\pi)$ is substantial and
consistent with a simple parabolic dispersion (with a gap at
$E_{\rm F}$). The asymmetric dispersion and the unclear Fermi
surface around $(\pi/2,\pi/2)$ for underdoped LSCO
are consistent with the electronic structures 
calculated by numerical exact diagonalization
on small clusters with stripes\cite{Tohyama}
and calculated within the Hubbard model
with the stripes using the Hartree-Fock
approximation\cite{Machida}
and dynamical mean-field theory.\cite{Fleck}

\begin{figure*}[!t]
	\epsfxsize=155mm
	\centerline{\epsfbox{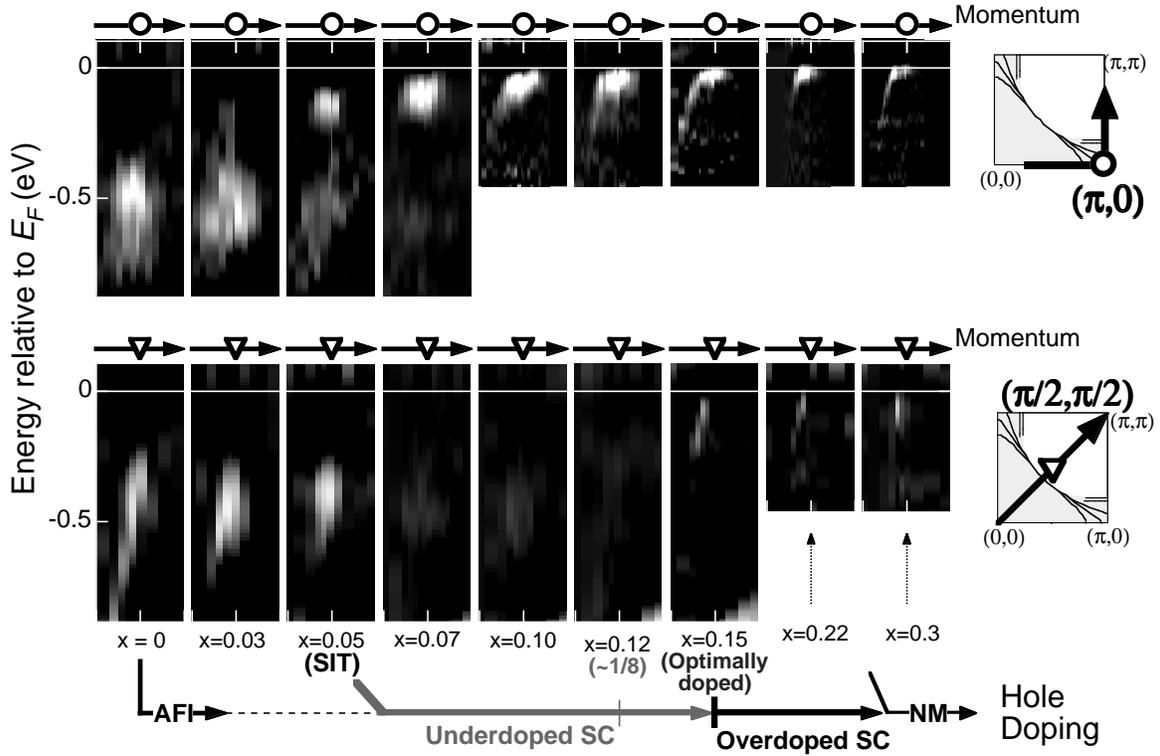}}
	\vspace{1pc} \caption{Doping dependence of the band
	dispersion around $(\pi,0)$ (upper panels) and
	$(\pi/2,\pi/2)$ (lower panels) for
	La$_{2-x}$Sr$_x$CuO$_4$.  From left to right, the
	samples are an antiferromagnetic insulator (AFI) for
	$x=0$, an insulator for $x=0.03$, near the
	superconductor-insulator transition (SIT) for
	$x=0.05$, superconductors (SC) for $x=0.07, 0.10,
	0.12, 0.15$ and 0.22, and a normal metal (NM) for
	$x=0.30$.  Data in the wide energy range were taken
	from Ref.~\protect\onlinecite{twocomponents}.  The
	features at $\sim-0.5$ eV in the underdoped samples
	are the lower Hubbard band.}
\label{disparound}
\end{figure*}


The band around $(\pi,0)$ is thought to primarily contribute to the 
formation of the superconducting condensate in a $d$-wave 
superconductor, because the quasiparticle weight near $E_{\rm F}$ 
around $(\pi/2,\pi/2)$ is virtually absent in underdoped LSCO 
($x\lesssim0.12$).  As shown in Fig.~\ref{flatband}, when LSCO is 
optimally doped, the flat band around $(\pi,0)$ is located slightly 
below $E_{\rm F}$ as in the other cuprate systems.  This is the case 
for all the hole-doped high-$T_c$ cuprates studied by ARPES so 
far,\cite{Marshall} suggesting that the energy position of the 
$(\pi,0)$ flat band has an universal doping dependence among 
high-$T_c$ cuprates, and that the optimum $T_c$ requires the $(\pi,0)$ 
flat band to be near $E_{\rm F}$.\cite{Imada} As for the relevance of 
the flat-band energy to the high-$T_{c}$, the presence of the flat 
band near $E_{\rm F}$ enhances the density of low-energy 
single-particle excitations which are involved in the formation of 
the superconducting condensate through a large portion of the 
${\bf k}$-space.\cite{Markiewicz}

In Fig.~\ref{disparound}, we summerize the doping
dependences of the dispersions around $(\pi,0)$ and
$(\pi/2,\pi/2)$.  It is clearly seen that the flat band
around $(\pi,0)$ is lowered as $x$ decreases and loses its
intensity in the insulating phase.  As reported
previously,\cite{twocomponents} the spectral weight is
transferred from the band near $E_{\rm F}$ ($\sim-0.1$ eV)
to the lower Hubbard band at $\sim-0.5$ eV in the vicinity
of the superconductor-insulator transition ($x\simeq0.05$). 
The evolution of the band near $E_F$ is different between
$(\pi,0)$ and $(\pi/2,\pi/2)$: with decreasing $x$, the
spectral weight is largely lost already at $x=0.12$ for
$\sim(\pi/2,\pi/2)$, whereas it remains substantial down to
$x=0.05$ for $\sim(\pi,0)$.  On the other hand, the
evolution of the insulating band at $\sim -0.5$ eV is
similar between $(\pi,0)$ and $(\pi/2,\pi/2)$.

\subsection{Fermi surface}

\begin{figure}
	\vspace{-1pt} \epsfxsize=41.5mm
	\centerline{\epsfbox{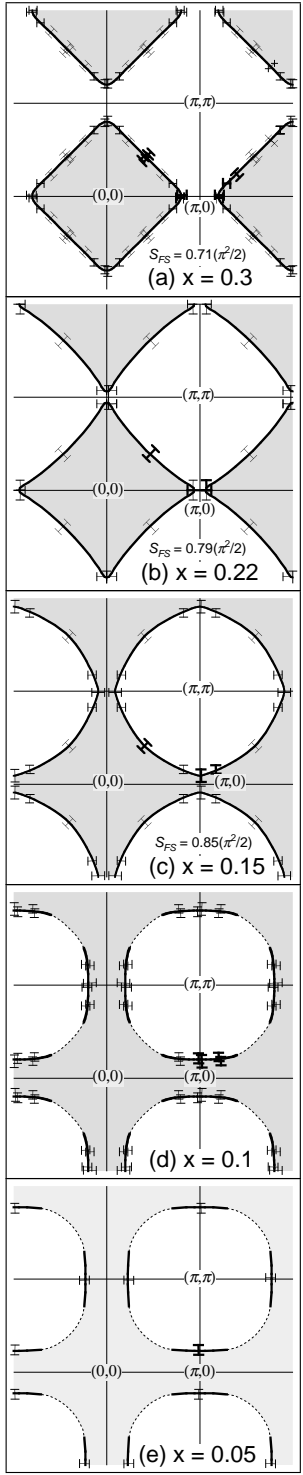}} 
	\vspace{-2pt} \caption{Fermi surfaces of
	La$_{2-x}$Sr$_x$CuO$_4$, obtained from ARPES
	experiments.  Thick and thin error bars denote the
	observed Fermi-surface crossings and those folded by
	symmetry.  As for $x=0.10$ and 0.05, since no
	dispersive features are observed near $E_{\rm F}$
	around $(\pi/2,\pi/2)$, the dotted curves are
	tentatively drawn so that the area enclosed by the
	Fermi surface is $\sim0.9$ and $\sim0.95$,
	respectively, of the half Brillouin zone area,
	assuming that the Luttinger sum rule is satisfied.}
\label{k-space}
\end{figure}

From the ARPES spectra taken at various doping levels, the
doping dependence of the Fermi surface has been deduced as
shown in Fig.~\ref{k-space}. Here the Fermi-surface crossings
have been determined to be the momenta where the
leading-edge energy reaches a local maximum and the spectral
peak intensity (quasiparticle weight) changes most strongly. 
They correspond to the minimum-gap loci, when a gap is
opened on the Fermi surface.  As for the superconducting
gap, it has been confirmed that the minimum-gap locus
coincides with the Fermi surface in the normal
state.\cite{Campuzano} In Fig.~\ref{k-space}, thick error
bars denote the actually measured positions of Fermi
surface, and the width of the error bars indicate two
momenta where the most weight of dispersive features is
clearly below $E_F$ and has almost gone above $E_F$.  The
area enclosed by the Fermi surface is $71\pm 3$, $79\pm 8$
and $85\pm 5\%$ of the half BZ area for $x=0.3$, 0.22 and
0.15, respectively, consistent with the Luttinger sum rule
for the electron density $1-x$ ($= 70, 78$ and $85\%$,
respectively).  As for $x=0.1$ and 0.05, since the Fermi
surface around $(\pi/2,\pi/2)$ was invisible, dotted curves
are tentatively drawn in Fig.~\ref{k-space} so that the area
enclosed by the Fermi surface is $\sim0.9$ and $\sim0.95$,
respectively,\cite{Tamasaku1} of the half BZ area, supposing
that the Luttinger sum rule is still satisfied.  As the hole
concentration decreases, the Fermi surface near $(\pi,0)$
smoothly moves through $(\pi,0)$ so that the topological
center of the Fermi surface is turned over from $(0,0)$ to
$(\pi,\pi)$ at $x\sim0.2$.  On the other hand, the position
of the Fermi surface near $(\pi/2,\pi/2)$ is less
sensitively dependent on the hole concentration and the weak
spectral intensity near $(\pi/2,\pi/2)$ at $E_{\rm F}$
becomes invisibly weak for $x\leq0.12$.\cite{twocomponents}
The Fermi surface of LSCO is thus strongly doping dependent,
while the Fermi surface of optimally doped LSCO is basically
similar to that of Bi2212.\cite{Marshall,DingDoping}

Figure~\ref{k-space} indicates that ``small hole pocket''
around $(\pi/2,\pi/2)$ is absent even in the underdoped
LSCO. Hence the decrease in the carrier density
proportional to $x$, which has been observed in the hall
coefficient measurement as $1/R_H \propto x$,\cite{Hall}
should be attributed to that the quasiparticle weight
around $E_{\rm F}$ decreases as $\propto x$ due to the
spectral weight transfer to higher binding
energies.\cite{twocomponents,AIPES}

\subsection{Energy gap}

\begin{figure}[!t]
	\epsfxsize=84mm \centerline{\epsfbox{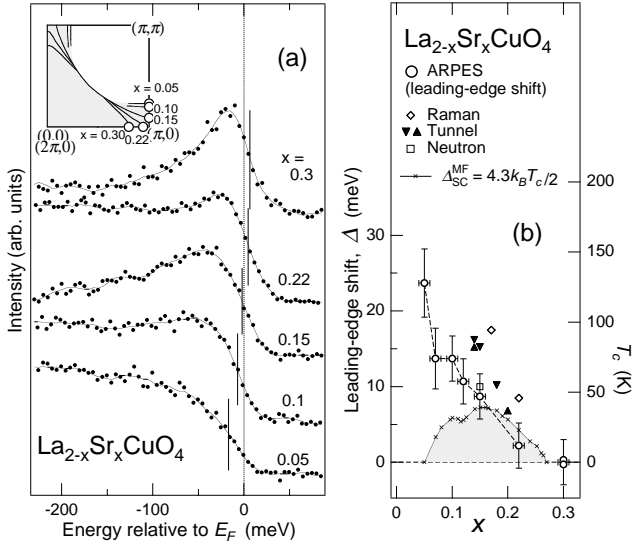}}
	\vspace{1pc} \caption{(a) ARPES spectra for momenta
	on the Fermi surface (minimum-gap locus) near
	$(\pi,0)$ as denoted by open circles in the inset. 
	From the ARPES spectrum for each composition $x$,
	the spectrum at $(0,0)$ has been subtracted as the
	angle-independent background.  (b) The shift
	$\Delta$ of the leading-edge midpoint in the ARPES
	spectra relative to that of $x=0.3$ ($\sim +6$ meV),
	denoted by open circles.  Error bars denotes the
	uncertainty in determining the leading-edge position
	of each spectrum.  The leading-edge shift $\Delta$
	approximately represents the magnitude of the
	superconducting or normal-state gap and is compared
	with the gap deduced from the $d$-wave mean-field
	approximation $2\Delta^{\rm MF}_{\rm SC} = 4.3k_{\rm
	B}T_{\rm c}$ (crosses) \cite{Won} and other
	experiments: Raman scattering (open
	diamonds),\cite{Raman} scanning tunnel spectroscopy
	(filled triangles),\cite{Tunnel} and inelastic
	neutron scattering (open boxes).\cite{INS} As $x$
	decreases, the magnitude of the energy gap keeps
	increasing even in the underdoped region in spite of
	the decreasing $T_{\rm c}$.}
	\label{scgap}
\end{figure}

The doping dependence of the energy gap at $E_{\rm F}$ may
be estimated from the leading-edge shift on the Fermi
surface.\cite{Ding-dgap,DingGap,PJWhite,Loeser,Harris,Norman,Harris-Bi2201,YBCO}
Figure~\ref{scgap}(a) shows the ARPES spectra at the momenta
where the leading edge reaches the maximum energy
(minimum-gap locus) around $(\pi,0)$ as shown by open
circles in the inset.  Here, the spectrum at $(0,0)$ has
been subtracted as the angle-independent background for each
composition.  For the non-superconducting ($x=0.3$) sample,
the leading-edge midpoint is apparently pushed above $E_{\rm
F}$ ($\sim 6$ meV) due to the finite instrumental resolution
($\sim 45$ meV).\cite{Loeser} As the hole concentration
decreases, the energy of the peak and the leading edge are
shifted downwards as a result of the opening of the
superconducting gap.

 In Fig.~\ref{scgap}(b), the energy shift $\Delta$ of the
leading-edge midpoint relative to that for $x=0.3$ ($\sim
+6$ meV) is plotted and compared with the results of other
experiments on LSCO, i.e., Raman scattering,\cite{Raman}
tunneling \cite{Tunnel} and neutron scattering\cite{INS}
studies (left axis).  Crosses indicate the superconducting
transition temperature $T_{\rm c}$ (right axis) and the
prediction of the mean-field theory for the $d$-wave
superconducting gap $2\Delta^{\rm MF}_{\rm SC}=4.3k_{\rm
B}T_{\rm c}$ \cite{Won} (left axis).  In fact, what are
measured in these experiments are different quantities,
e.g., the neutron scattering measures the gap in the
spin-excitation spectrum, which is not simply connected to
the single particle excitation gap probed by ARPES. In
addition, the magnitude of the ARPES leading-edge shift
tends to be smaller than the tunneling result, probably
because the broadness of the peak reduces the apparent
shift of the ARPES leading edge, while it hardly affects
the peak position observed in  tunneling spectra, which
represent the momentum-integrated spectral function. 
Nevertheless the doping dependence of the gap magnitude is
consistent among the ARPES and the other experiments.

\begin{figure}[!t]
	\epsfxsize=57mm
	\centerline{\epsfbox{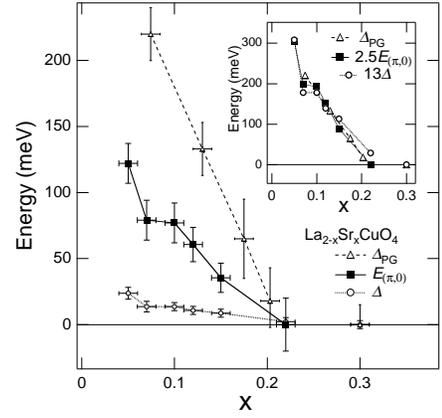}}
	\vspace{0.5pc} \caption{Doping dependence of three
	characteristic energies: the leading-edge shift on
	the Fermi surface $\Delta$ representing the
	superconducting and normal-state gaps, the energy of
	the flat band around $(\pi,0)$, $E_{(\pi,0)}$, and
	the ``large pseudogap'' $\Delta_{\rm PG}$, which
	would correspond to the high energy bump in
	angle-integrated photoemission (AIPES)
	spectra.\cite{AIPES,TT-LSCO} Error bars for
	$E_{(\pi,0)}$ indicate the uncertainties in
	determining the peak energy at
	$\sim(\pi,0)$\cite{Epi0}, based on variation among
	several different samples.  Inset shows the scaling
	relations as $\Delta_{\rm PG} \approx 2.5
	E_{(\pi,0)}$ and $\Delta_{\rm PG} \approx
	13\Delta$.}
\label{scaling}
\end{figure}

As the hole concentration $x$ decreases, the magnitude of
$\Delta$ keeps increasing even in the underdoped region, in
spite of the decreasing $T_{\rm c}$.  This remarkable
feature has also been reported for Bi2212
\cite{PJWhite,Harris} and is thus likely to be an universal
feature of the cuprate superconductors.  The present data
have ensured that this tendency is sustained down to
$x=0.05$.  Although the sample of $x=0.05$ is not
superconducting, still an energy gap is opened at
$\sim(\pi,0.25\pi)$ as shown in Fig.~\ref{x=0.05}(b),
corresponding to the ``normal-state gap'' observed for
underdoped
Bi2212.\cite{DingGap,PJWhite,Loeser,Harris,Norman} From the
ARPES spectra [Figs.~\ref{x=0.05}(b) and \ref{scgap}(b)], it
appears that the superconducting gap smoothly evolves into
the normal-state gap with decreasing hole concentration $x$. 
This observation certainly has the same significance as the
fact that the temperature dependence of the leading-edge
shift is continuous at $T_{c}$ for underdoped
Bi2212,\cite{DingGap,PJWhite,Loeser,Harris,Norman} These
connections between the normal-state and superconducting
gaps suggest that these gaps have the same origin.  Assuming
that the magnitude of the energy gap $\Delta$ represents the
paring strength, the doping dependence of $T_c$ may be
roughly described using the product of $\Delta$ and the
quasiparticle density at $E_F$ related to the flat-band
energy.  When the hole concentration is further decreased
to $x < 0.05$, the normal-state gap becomes difficult to
be identified because the spectral weight of the band near
$E_{\rm F}$ diminishes, and alternatively the wide
insulating gap ($\sim0.5$ eV) becomes
predominant\cite{twocomponents}.

Figure~\ref{scaling} shows the binding energy $E_{(\pi,0)}$
of the band at $(\pi,0)$\cite{Epi0} which is confidently
determined by measuring several samples for each $x$,
compared with the energies of the superconducting or
normal-state gap $\Delta$ measured by the ARPES leading-edge
shift and the ``large pseudogap'' $\Delta_{\rm PG}$, which
would correspond to the high energy bump in AIPES
spectra.\cite{AIPES,TT-LSCO} These characteristic energies
show quite similar doping dependences as shown in the inset,
even though their energy scales are different: $\Delta_{\rm
PG} \approx 2.5 E_{(\pi,0)}$ and $\Delta_{\rm PG} \approx
13\Delta$.  Therefore, the electronic structure of the
underdoped LSCO is essentially characterized by a single
parameter which rapidly increases as $x$ decreases for $x
\lesssim 0.22$.  The proportionality $ \Delta_{\rm PG}
\approx 2.5 E_{(\pi,0)} \approx 13\Delta$ implies that the
origin of the superconducting and normal-state gaps may be
related to that of the large pseudogap and the flat-band
energy, indicating that the behaviors of the cuprate
superconductors are strongly affected by the short-range
antiferromagnetic
correlations.\cite{AIPES,Shen&Schrieffer,Nakano}

\subsection{Self-energy analysis}
\label{self}

In order to deduce the energy position and width of the ARPES peak 
more precisely, a model for the spectral lineshape is necessary.  The 
actual peak is asymmetric and fairly deviated from the simple 
Lorentzian even for the heavily overdoped sample ($x=0.3$) 
\cite{FermiSurface}.  Therefore, we introduce a simple but more 
general form of the self-energy\cite{SaitohSelf}: $$\Sigma(\omega) = - 
\frac{\gamma}{\omega/\Gamma+i} + \frac{\gamma+\gamma_0}{\omega/G+i} 
\qquad\quad (G\gg\Gamma),$$ which satisfies Kramers-Kronig relation.  
The denominator of the second term is to make $\Sigma(\omega)$ 
converge to zero for $\omega \rightarrow \infty$, a sufficiently large 
$G$ being taken as a cut-off energy.  Then, for $\omega\ll G$, 
$\Sigma(\omega)$ is expanded around $E_F$ as $\Sigma(\omega) \sim 
-\gamma(\omega/\Gamma) -i\gamma_0 -i\gamma(\omega/\Gamma)^2$.  Here, 
$\Gamma$ is the characteristic energy which scales for the 
quasiparticle energy $\omega$, $\gamma_0 = -{\mathrm Im}\Sigma(0)$ 
represents the scattering rate of the quasiparticles at $\omega=0$ and 
should be zero for an ideal Fermi-liquid, and $\gamma$ represents the 
high energy limit of the peak width since $-{\mathrm Im}\Sigma(\omega) 
\simeq \gamma+\gamma_0$ for $\Gamma \ll \omega \ll G$.  In the present 
analysis, the momentum dependence of the self-energy is ignored for 
the simplicity.  Then, the spectral function $A({\mathbf k},\omega)$ 
is given by $$A({\mathbf k},\omega) = {\mathrm 
Im}\left(\frac{1}{\omega -
\epsilon_{\mathbf k} - \Sigma({\mathbf k},\omega)}\right),$$
where $\epsilon_{\mathbf k}$ is the dispersion of the single particle 
band.  The calculated spectra have been obtained as the product of 
$A({\mathbf k},\omega)$ and the Fermi-Dirac distribution function 
$f(\omegaÁ,T)$, and then broadened by the energy and angular 
resolutions (42 meV and 2$^\circ$, respectively).  Finally, upon 
comparing with experimental spectra, the angle-independent background, 
i.e., the spectrum at (0,0), is commonly added to the calculated 
spectra.

Parameters fixed in the analysis are the temperature and the energy 
and momentum resolutions, and the single-particle dispersion 
$\epsilon_{\mathbf k}$ has been taken from the 
local-density-approximation (LDA) energy band of undoped 
La$_2$CuO$_4$\cite{bandcalcLSCO}.  On the other hand, the parameters, 
$\Gamma$, $\gamma$ and $\gamma_0$, describing the self-energy are 
obtained from the present least-square-fit analysis, and the results 
are shown in Table~\ref{parameters}.  Here, the chemical potential 
shift of the LDA band due to the hole doping into La$_2$CuO$_4$ is 
adjusted to reproduce the experiment, and the spectral intensity at 
each angle has also been adjusted to the experiment, because the 
momentum dependence of the matrix element is unknown.

\begin{figure}[!t]
	\epsfxsize=84mm
	\centerline{\epsfbox{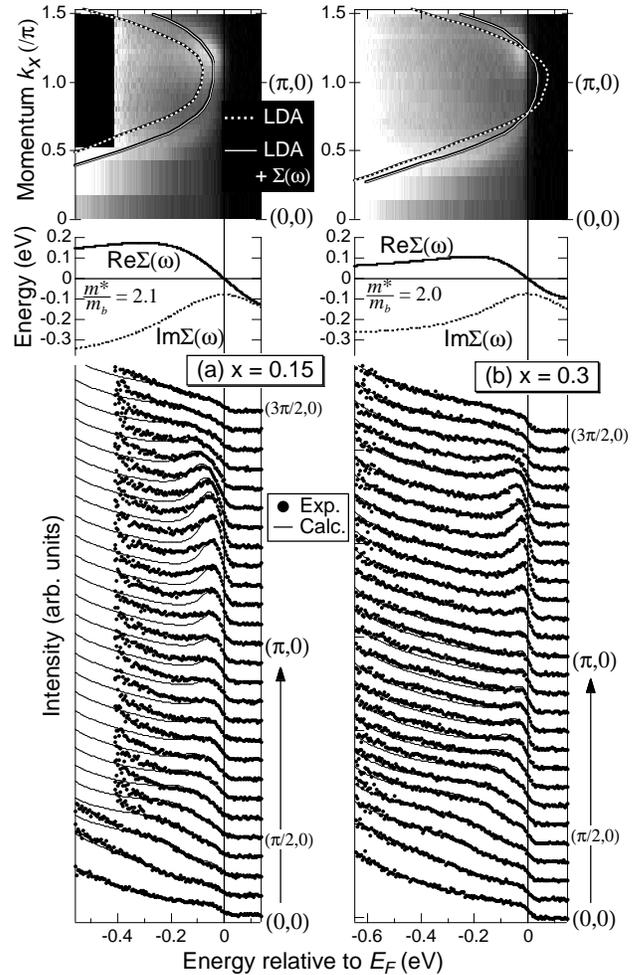}}
	\vspace{0.5pc} \caption{Results of the self-energy
	analysis for (a) $x=0.15$ and (b) $x=0.30$.  Bottom
	panels: calculated spectra (lines) fitted to the
	experimental spectra (dots) along
	$(0,0)\rightarrow(\pi,0)$.  Middle panels: real and
	imaginary parts of the self-energy used in the
	calculation.  Top panels: dispersion
	$\epsilon_{\mathrm k}$ of the
	local-density-approximation energy band of
	La$_2$CuO$_4$ whose chemical potential adjusted to
	experiment (dotted lines), and the peak dispersion
	in the calculated spectra (solid lines), overlayed
	with the gray scale plot of the experimental
	spectra, where the white region simply denotes high
	spectral intensity (not the second derivatives).}
\label{selfenergy}
\end{figure}

\begin{table*}[!t]
\caption{Effective mass enhancement factor $m^*/m_b$ at $E_F$ obtained 
from the self-energy analysis of the ARPES spectra for $x=0.30$ (overdoped)
and $x=0.15$ (optimally doped), and parameters $\Gamma$, $\gamma$ and $\gamma_0$
of the model self-energy which best reproduce the experimental spectra.
Here $m^*/m_b$ obtained from the electronic specific heat
coefficient $\gamma_{el}$\cite{Momono} are also shown for comparison.}
\begin{center}
\begin{tabular}{ccccccc} 
$x$ & $m^*/m_b$ (ARPES) & $\Gamma$ (eV) & $\gamma$ (eV) & $\gamma_0$ (eV) & Fermi surface & $m^*/m_b$ ($\gamma_{el}$) \\ \hline
$0.30$ & 2.0 & 0.20 & 0.21 & 0.077 &  centered at $(0,0)$ & 2.5\\
$0.15$ & 2.1 & 0.30 & 0.34 & 0.081 &  centered at $(\pi,\pi)$ & 2.5 \\
\end{tabular}
\end{center}
\label{parameters}
\end{table*}

Figure~\ref{selfenergy} shows the results of such analysis for 
$(0,0)\rightarrow(\pi,0)$ cut.  Both the peak lineshape and the peak 
dispersion are successfully reproduced for the heavily overdoped 
sample ($x=0.30$), confirming the dispersion relation and Fermi 
surface crossings shown in Figs.~\ref{flatband} and \ref{k-space}.  
Note that the weak residual spectral weight around $(\pi,0)$ is also 
present in the calculated spectra even though the band energy 
$\epsilon_{\mathbf k}$ at $(\pi,0)$ is above $E_F$, indicating that 
the experimental spectra are consistent with the Fermi surface 
centered at $(0,0)$.  For the optimally doped sample ($x=0.15$), on 
the other hand, the high-energy tail of the peak was difficult to 
reproduce after extensive trials particularly around $(\pi,0)$, 
although the peak dispersion and the peak leading-edge are almost 
correctly reproduced by a self-energy similar to that of $x=0.30$.  
The result for $x=0.15$ indicates that the high-energy tail around 
$(\pi,0)$ contains an intense incoherent component which cannot be 
described by the simple model self-energy analysis.  This difficulty, 
in addition to the limited experimental resolution, indicate that the 
obtained ${\mathrm Im}\Sigma(\omega)$, which describes the peak shape, 
has some uncertainties, while ${\mathrm Re}\Sigma(\omega)$ obtained 
from the peak position is reliably determined.

The effective mass $m^*$ relative to the bare-electron mass
$m_b$ is also obtained from the self energy:
$$\frac{m^*}{m_b} = 1 - \left.\frac{\partial{\mathrm
Re}\Sigma(\omega)}{\partial\omega}\right|_{\omega=0} = 1 +
\frac{\gamma}{\Gamma},$$ 
and the result is shown in Table~\ref{parameters}.  Namely, the 
electron effective mass $m^*$ has been directly obtained from the band 
dispersion around the Fermi level.  The effective-mass enhancement 
factor $m^*/m_b$ deduced from ARPES spectra is approximately 
consistent with that from the electronic specific heat coefficients 
$\gamma_{el}$ \cite{Momono}, indicating that the used self-energy is 
reasonable to some extent.  Furthermore, we also find the peak in the 
spectral function at $E_F$ has a width of $2\gamma_0(m_b/m^*) = 77$ 
meV for both $x=0.30$ and 0.15, in addition to the broadening due to 
the instrumental resolutions.  The quantities of $-{\mathrm 
Im}\Sigma(0) = \gamma_0$ obtained by the present analysis are 
approximately consistent with that the result for overdoped Bi2212 is 
$\sim90$ meV independent of temperature in the normal state $T>T_c$ 
\cite{BSCCOself}.  Although the spectrum of the $x=0.15$ sample was 
taken in the superconducting state, our preliminary 
temperature-dependent measurements indicated no significant broadening 
of the peak above $T_c$ except for the thermal broadening.

\section{Conclusions}

In summary, the systematic ARPES study of LSCO has revealed
the evolution of the Fermi surface, the superconducting gap
and the band dispersion around $(\pi,0)$ with hole doping. 
While the Fermi surface and the band dispersion of the
optimally doped LSCO are essentially consistent to the
result of Bi2212,\cite{Marshall} those low-energy electronic
structures have been found to change drastically for the
wide hole concentration range ($0.05<x<0.30$) available for
LSCO. Notably, the magnitude of the superconducting gap
$\Delta$ keeps increasing as $x$ decreases down to $x=0.05$,
and the superconducting gap appears to evolve smoothly into
the normal-state gap for $x=0.05$.  It has been shown
that the doping dependence of $\Delta$ deviates from the
decreasing $T_c$ in the underdoped region but follows a
doping dependence common to other two characteristic
energies: the energy $E_{(\pi,0)}$ of the extended flat band
at $\sim(\pi,0)$ and the pseudogap energy $\Delta_{PG}$
obtained from AIPES. Therefore, the electronic structure of
the underdoped cuprates may be characterized by a single
parameter.  For the heavily overdoped region ($x=0.30$), the
simple self-energy analysis have successfully reproduced
both the band dispersion and the spectral lineshape and
indicated the effective mass $m^*/m_b \sim 2$.  However, as
hole concentration decreases, the incoherent component which
cannot be described by the simple self-energy analysis
grows intense in the high-energy tail of the ARPES peak.  As
the flat band at $(\pi,0)$ is lowered with decreasing $x$,
the band dispersion along $(\pi,0)\rightarrow(\pi,0.3\pi)$
becomes faster, while almost no dispersion along
$(\pi,0)\rightarrow(0.7\pi,0)$ is kept. Such electronic
structure is consistent with some stripe-model calculations.
This picture is also supported by the earlier
observation of two components in
the electronic structure.\cite{twocomponents}

\section*{Acknowledgment}

This work was supported by the New Energy and Industrial
Technology Development Organization (NEDO), a Special
Coordination Fund for Promoting Science and Technology from
the Science and Technology Agency of Japan, a Grant-in-Aid
for Scientific Research ``Novel Quantum Phenomena in
Transition Metal Oxides'' from the Ministry of Education,
Science, Culture and Sports of Japan, and the U.~S.~DOE, Office
of Basic Energy Science and Division of Material Science. 
Stanford Synchrotron Radiation Laboratory is operated by the
U.~S.~DOE, Office of Basic Energy Sciences, Division of
Chemical Sciences.


\begin{references}
\bibitem{broadening}
If the peak at $E_F$ in the spectral
function were a delta function as expected for a
conventional Fermi liquid, the finite resolution would make
no effect on the peak position.  However, the actual ARPES
spectra seem to have a finite peak width ($\sim77$ meV) at
$E_F$ in addition to the broadening due to the energy and
angular resolutions, as indicated by the analysis in
Sec.~\ref{self}.  Therefore, the ARPES peak due to a band
near $E_F$ appears below $\sim-\Delta E/2$ in the acutual
spectra.
\bibitem{Epi0}
As for $x=0.22$, it is due to the Fermi cut-off and the
finite instrumental resolution that a weak spectral peak
appears slightly below $E_F$ at $(\pi,0)$. 
In reality, the Fermi-surface crossing for $x=0.22$ is quite
close to the $(\pi,0)$ point as discussed in Sec. III. A and C.
Therefore, in Fig.~\ref{scaling} (b), $E_{(\pi,0)}$ for $x=0.22$
is plotted at zero with a relatively large error bar.
\end{references}
\end{document}